\documentclass[a4paper,11pt]{article}
\pdfoutput=1 

\usepackage{jcappub}

\usepackage[utf8]{inputenc}
\usepackage{graphicx}
\usepackage{mathtools}
\usepackage{amsmath}
\usepackage{wrapfig}
\usepackage{sidecap}
\usepackage{fullpage}
\usepackage{multirow}
\usepackage{siunitx}
\usepackage{listings}
\usepackage{cancel}
\usepackage{braket}
\usepackage{xcolor}
\usepackage{gensymb}

\title{Neutrino emission upper limits with maximum likelihood estimators for joint astrophysical neutrino searches with large sky localizations}
\author[a,1]{Do\u{g}a Veske,\note{Corresponding author}}
\author[b]{Zsuzsa M\'arka,}
\author[c]{Imre Bartos}
\author[a]{and Szabolcs M\'arka}

\affiliation[a]{Department of Physics, Columbia University in the City of New York, 550 W 120th St., New York, NY 10027, USA}
\affiliation[b]{Columbia Astrophysics Laboratory, Columbia University in the City of New York, 550 W 120th St., New York, NY 10027, USA}
\affiliation[c]{Department of Physics, University of Florida, PO Box 118440, Gainesville, FL 32611-8440, USA}

\emailAdd{dv2397@columbia.edu}
\emailAdd{zsuzsa@astro.columbia.edu}
\emailAdd{imrebartos@ufl.edu}
\emailAdd{sm2375@columbia.edu}

\abstract{
Since the start of the gravitational wave observation era, no joint high energy neutrino and gravitational wave event has been found. These non-detections could be used for setting an upper bound on the neutrino emission properties for gravitational wave events individually or for a set of them. Although in the previous joint high energy neutrino and gravitational wave event searches upper limits have been found, there is a lack of consistent method for the calculation. The problem addressed in this paper is finding those limits for astrophysical events which are localized poorly in the sky where the sensitivities of the neutrino detectors change significantly and can also emit neutrinos, for example the gravitational wave detections. Here we describe methods for assigning limits for expected neutrino count, emission fluence and isotropically equivalent emission based on maximum likelihood estimators. Then we apply described methods on the three GW detections from aLIGO's first observing run (O1) and find upper limits for them.}
\begin{document}
\maketitle
\flushbottom

\section{Introduction}
With the observational discovery of gravitational waves (GW), humanity acquired another way of observing the universe \cite{PhysRevLett.116.061102}. It could allow us to observe phenomena which we are not able to see with other methods such as observations based on electromagnetic or neutrino emission, as well as observing cosmic events which could be observed via a multitude of messengers~\cite{2008CQGra..25k4051A,2013CQGra..30l3001B}. There has been three such fundamental observations so far; the binary neutron star (BNS) merger which was observed with GWs and electromagnetic radiation in various bands~\cite{Abbott_2017}, the blazar which was observed with high energy neutrinos (HEN) and electromagnetic radiation~\cite{blazar} and the SN1987a supernova which was observed with lower energy neutrinos (in MeV energy) and in various electromagnetic bands \cite{doi:10.1146/annurev.aa.27.090189.003213}. Having one or more additional messengers beyond GWs might shed a brighter light on the physical processes happened before, during, and after the astrophysical event \cite{M_sz_ros_2019,METZGER2019167923} which can enable 
having new observations that wouldn't be possible \cite{BartosMarkaNature2019,2019ApJ...881L...4B}. Moreover absence of an additional messenger is also informative as in the case of the first multimessenger observational result with GWs that addressed the origin of a short-hard gamma-ray burst from the direction of the Andromeda galaxy~\cite{2008ApJ...681.1419A}. Despite well over a decade long effort ~\cite{2008CQGra..25k4039A,2009IJMPD..18.1655V,2011PhRvL.107y1101B,PhysRevD.85.103004,2013JCAP...06..008A,2013RvMP...85.1401A,2014PhRvD..90j2002A,PhysRevD.93.122010,Albert_2017,PhysRevD.96.022005,Albert_2019}, one remaining two messenger combination is a joint observation with HENs and GWs.

During aLIGO's first observing run (O1) and second observing run with aVirgo (O2) searhces for joint GW and HEN events couldn't find any significant event \cite{Albert_2019, PhysRevD.93.122010, PhysRevD.96.022005, Albert_2017}. Moreover during the first half of the third observing run of aLIGO and aVirgo (O3a), search on every public alert of LIGO-Virgo Collaboration is performed \cite{gracedb}; no significant event was found and the results for each search was reported via The Gamma-ray Coordinates Network (GCN) Notices by IceCube Collaboration \cite{keivani2019multimessenger,hussain2019search,gcn} Such searches are based on assigning a test statistic (TS) to each observed HEN, which are detected in a fixed time window before and after the detected GW \cite{PhysRevD.85.103004, PhysRevD.100.083017}. Based on the value of the TS for each detected event, a significance is assigned to each event by comparing it to a known background TS distribution. Unless the observed event's TS exceeds a threshold, which is determined by a fixed significance level and the background TS distribution, the event is not counted as a multi-messenger detection. One physical information that can be extracted from GW detections is an upper limit on the neutrino emission fluence (time and energy integrated flux in units of energy per area), on the expected neutrino count or an isotropically equivalent neutrino emission energy from the GW event's source. In case of a non-detection these upper limits can still be used to make inferences. Although in the previous searches upper limits were calculated for these quantities in various ways, there hasn't been a consistent method used for those calculations \cite{Albert_2019, PhysRevD.93.122010, PhysRevD.96.022005}. In \cite{PhysRevD.93.122010, PhysRevD.96.022005}, upper limits on emission fluence for each point in the sky, and lowest and highest upper limits on isotropically emitted energy for 90 and 95\% credible levels of localization are given for individual events. In \cite{Albert_2019} a non-explained count of 3.9 events is used for determining a rate upper limit. In none of them a single limit on fluence or energy for each event is not found which requires proper handling of the direction dependent sensitivities of the detectors together with events' large localizations. 

The neutrino detectors’ sensitivities are usually direction dependent, therefore the required fluence/energy for a given number of event number depends on the sky location of the neutrino. When the localization of the event is large compared to the directional change of the sensitivity (i.e. if the sensitivity varies significantly in the localization region as in the case for GW localizations and neutrino detectors) then it would not be trivial to find a single fluence/energy value corresponding to a given neutrino count. In previous studies, instead of a single limit for the events, limits for every point in the localization were found. Due to the need of having such a formalism, with this paper we propose a method for calculating frequentist upper limits based on maximum likelihood estimators. 

Although the TS can directly be used for finding an upper limit without an estimator for a higher statistical power, it can result in extremely low upper limits or empty confidence intervals when less background events are observed than the expectation. One of the \emph{ad-hoc} solutions currently used for this problem is to put a lower bound to TS equal to the median of the background TS distribution, namely the sensitivity, when calculating upper limits directly from TS. Moreover due to possible nonideal constructions of TS (such as approximations and assumptions), unphysical small variations can be present which can unnecessarily differentiate between the physically identical observations. These unphysical variations can be systematic variations at any part of the analysis which are not accounted for completely. Limitations on the energy estimation of individual neutrinos or simplifying assumptions in the TS which do not represent the actual complicated physics can be given as examples for such variations. By the Neyman's construction of frequentist upper limits \cite{doi:10.1098/rsta.1937.0005}, these variations reduce (tighten) the limits unnecessarily. The method we present here brings solution to both of these problems. It restores the unphysical variations by using the distributions calculated from physical simulations and finds upper limits about the background-free upper limits (i.e. 2.3 for 90\% upper limits) even when highly non-significant observation is made (i.e. less number of background is seen than expectation), as we demonstrate at the end of the paper with real events.

As the sensitivities of GW observatories get better and number of GW observatories in the GW detection network increases, the rate of GW observations will increase. For example, in the first 6 months of the third observing run of aLIGO and aVirgo 33 GW candidate detections have been announced publicly compared to total 11 detections announced from O1 and O2 after offline analysis \cite{Abbott_2019}. Thus, it is expected to have a population for the GW detections in future. Therefore a proper quantification of the upper limits for the counterparts of these GW events will let us infer about the physics involved in them. 

In this paper we describe a method of finding frequentist upper limits based on maximum likelihood estimators for expected neutrino count and neutrino emission fluence for single GW events which we explain in Sections \ref{count} and \ref{fluence} respectively; and in Section \ref{sec:eiso} we describe the method for finding the upper limit on the isotropically equivalent emitted energy for an ensemble of events of same kind, assuming the energy is same for all. Considering the correspondence of frequentist limit from maximum likelihood estimation and Bayesian limit for uniform prior on an ideally counted Poisson variable, we also compare frequentist and Bayesian limits for uniform prior for the neutrino count which is inferred by a TS in Sec. \ref{sec:bayescomp}. Finally in Section \ref{sec:res} we demonstrate our method by finding the upper limits for the 3 GW events detected during the first observing run of aLIGO (O1).

Generally a 90\% upper limit is desired to be found in the joint GW-HEN event searches \cite{Albert_2019, PhysRevD.93.122010, PhysRevD.96.022005} although it is an arbitrary confidence interval. Without loss of generality, throughout this paper we aim to find 90\% upper limits for clarity. However any kind of confidence interval at any confidence level can be found with this method, by requiring a different relationship between the estimators of the true quantity and the measurement at the very end of the calculation. Furthermore, due to inferred low chance of detecting a joint HEN with GW from absence of such an example, again without loss of generality we also assume that we are dealing with 2 or less neutrinos detected from the same source with the GW for simplicity. We will refer such neutrinos as \emph{signal} neutrinos, and all other detected neutrinos, which may be real or noise originated and are not associated with GWs, as \emph{background} neutrinos.
\section{Upper limit for neutrino count}
\label{count}
\subsection{Frequentist upper limit}
In this section we find the 90\% upper limit for the expected number of neutrinos from a single measurement via maximum likelihood estimators, where the measurement's significance is quantified by a test statistic (TS). In order to write the likelihood we need to have the TS distributions for 0, 1 and 2 signal neutrinos.
First write the likelihood and denote the expected number of neutrinos with \(\theta\);
\begin{equation} \mathcal{L}(\theta;TS_m)=P(TS_m|\theta)=\sum_{n=0}^\infty P(TS_m|\theta,n)P(n|\theta),\ \theta\geq 0\end{equation} 
where \(n\) is the number of signal neutrinos from the common source and \(TS_m\) is the measured TS. As said before we assume to have at most 2 signal neutrinos.
\begin{equation}
\mathcal{L}(\theta;TS_m)=  e^{-\theta}P_0(TS_m)+\theta e^{-\theta}P_1(TS_m)+\frac{\theta^2}{2} e^{-\theta}P_2(TS_m)  \end{equation}
where \(P_0,\ P_1,\ P_2\) are the TS distributions corresponding to 0, 1, and 2 signal neutrinos.
Now find the local maximum of the likelihood. After taking derivative with respect to \(\theta\) and equating to 0, we have
\begin{equation}
P_1(TS_m)-P_0(TS_m)   \\ + (P_2(TS_m)-P_1(TS_m))\theta-P_2(TS_m)\frac{\theta^2}{2}  =0 \end{equation}
we find the local maximum at 
\begin{equation} \hat{\theta}=1-\frac{P_1(TS_m)}{P_2(TS_m)}+\sqrt{1+\frac{P_1(TS_m)^2}{P_2(TS_m)^2}-\frac{2P_0(TS_m)}{P_2(TS_m)}}\end{equation} 
(other root corresponds to local minimum). For \(P_2(TS_m)=0\) it is 
\begin{equation} \hat{\theta}=1-\frac{P_0(TS_m)}{P_1(TS_m)}\end{equation} 
For \(P_2(TS_m)=P_1(TS_m)=0\) there is no local maximum. The absolute maximum is at, \begin{equation} \hat{\theta}=0\end{equation} 
The hat on \(\theta\) is for denoting the maximum likelihood estimator. Local maximum may not be the absolute maximum in \(\theta=[0,\infty)\). The only other candidate for absolute maximum is the border value \(\hat{\theta}=0\). 
So maximum likelihood estimator for expected neutrino number is either 0 or \(1-\frac{P_1(TS_m)}{P_2(TS_m)}+\sqrt{1+\frac{P_1(TS_m)^2}{P_2(TS_m)^2}-\frac{2P_0(TS_m)}{P_2(TS_m)}}\), it can be determined by comparing the value of likelihood at these two points. Let's keep denoting it as \(\hat{\theta}\).\\
Now the upper limit for expected neutrino number is defined as the neutrino number above which we have 90\% probability to have a higher maximum likelihood estimator than the one for our measurement, in mathematical notation we find the \(\theta\) which satisfies the Equation \eqref{theta}
\begin{equation} \label{theta} P(\hat{\theta}<\hat{\theta}'|\theta)=0.9\end{equation} 
where \(\hat{\theta}'\) is the maximum likelihood estimator for \(\theta\), and \(\hat{\theta}\) is the maximum likelihood estimator for our current measurement. It is equivalent to 
\begin{equation} \int_{0^-}^{\hat{\theta}+0^{+}} f(\hat{\theta}'|\theta)d\hat{\theta}'=0.1 \label{eq:freqcdf}\end{equation} 
where \(f(\hat{\theta}'|\theta)\) is the probability distribution function for the maximum likelihood estimator for true expected neutrino number \(\theta\). It consists of delta distributions at integers and a continuous distribution between \(\hat{\theta}'=[0,2]\). Delta distribution at integers \(\geq 3\) are identical to Poisson distribution and strengths' of delta distributions at 0, 1 and 2 are less than the corresponding Poisson strengths. The missing probability is contained in the continuous distribution.
\subsection{Bayesian limit with uniform prior}

Now we discuss the Bayesian credible limits with for a uniform prior \(\theta\). Frequentist limits with maximum likelihood estimators and Bayesian limits for uniform priors give the same limits if the quantity of interest is a location parameter \cite{Jaynes1976}, such as the mean of a Gaussian distribution. Moreover as shown in section \ref{ideal}, they also correspond to each other for the mean of a Poisson distribution although the mean of Poisson distribution is not a location parameter.

Bayesian upper limit requires having a posterior distribution for \(\theta\) as
\begin{equation} P(\theta|TS_m)=\frac{P(TS_m|\theta)P(\theta)}{P(TS_m)}\end{equation} 
where \(P(TS_m)\) acts like a normalization constant. For uniform prior \(P(\theta)=k\)
\begin{equation} P(\theta|TS_m)=\frac{kP(TS_m|\theta)}{P(TS_m)}\end{equation} 
Again if we are sure that signal neutrino number \(\leq 2\)
\begin{equation}P(\theta|TS_m)= \\ \frac{e^{-\theta}P_0(TS_m)+\theta e^{-\theta}P_1(TS_m)+\frac{\theta^2}{2} e^{-\theta}P_2(TS_m)}{P_0(TS_m)+P_1(TS_m)+P_2(TS_m)} \end{equation}
The Bayesian 90\% upper limit is defined as the point where cumulative posterior probability is 0.9, or  the \(\theta'\) which satisfies the Equation \eqref{bayes}
\begin{equation} P(\theta<\theta'|TS_m)=\int_0^{\theta'} P(\theta|TS_m) d\theta = 0.9 \label{bayes}\end{equation} 
Hence
\begin{equation} e^{-\theta'}(1+\frac{\theta'(P_1(TS_m)+P_2(TS_m))+\frac{\theta'^2}{2}P_2(TS_m)}{P_0(TS_m)+P_1(TS_m)+P_2(TS_m)})=0.1 \label{eq:bayesfinal}\end{equation} 
\subsection{Comparison of frequentist and Bayesian limits for uniform prior}
Now we compare the frequentist and Bayesian limits for a uniform prior for the quantity we estimate. First we show that these two limits have correspondence for the ideal Poisson counting case and then we show that when we have a measurement with a TS they do not necessarily correspond to each other.
\label{sec:bayescomp}
\subsubsection{Ideal Poisson counting experiment}
\label{ideal}
Before comparing frequentist and Bayesian limits for uniform prior for the case of interest, compare them for a Poisson counting experiment where we count the events and the maximum likelihood estimator and the TS are equivalent to the observed event count, in other words there is no ambiguity in the signal neutrino count. Denote it with \(\theta_{obs}\). The frequentist 90\% upper limit \(\theta_L\) satisfies
\begin{equation} \sum_{n=\theta_{obs}+1}^{\infty} Poisson(n,\theta_L)=\sum_{n=\theta_{obs}+1}^{\infty} \frac{\theta_L^n e^{-\theta_L}}{n!}=0.9\end{equation} 
The Bayesian posterior distribution with uniform prior becomes
\begin{equation} P(\theta|\theta_{obs})=Poisson(\theta_{obs},\theta)\end{equation} 
and the Bayesian limit \(\theta_L\) satisfies
\begin{equation} \int_0^{\theta_L} \frac{\theta^{\theta_{obs}}e^{-\theta}}{\theta_{obs}!} d\theta= 0.9\end{equation} 
Now compare \( \int_0^{\theta_L} \frac{\theta^{\theta_{obs}}e^{-\theta}}{\theta_{obs}!} d\theta\) and \(\sum_{n=\theta_{obs}+1}^{\infty} \frac{\theta_L^n e^{-\theta_L}}{n!}\). First differentiate them with respect to \(\theta_L\). From the sum's derivative we have
\begin{equation} \sum_{n=\theta_{obs}+1}^{\infty} \frac{\theta_L^{n-1} e^{-\theta_L}}{(n-1)!}-\frac{\theta_L^n e^{-\theta_L}}{n!}=\frac{\theta_L^{\theta_{obs}} e^{-\theta_L}}{\theta_{obs}!}\end{equation} 
From the integral's derivative we have
\begin{equation} \frac{\theta_L^{\theta_{obs}}e^{-\theta_L}}{\theta_{obs}!}\end{equation} 
the same expression. Hence \( \int_0^{\theta_L} \frac{\theta^{\theta_{obs}}e^{-\theta}}{\theta_{obs}!} d\theta\) and \(\sum_{n=\theta_{obs}+1}^{\infty} \frac{\theta_L^n e^{-\theta_L}}{n!}\) can only differ by a constant. However when \(\theta_L=0\) they are both 0. Therefore equations
\begin{equation} \int_0^{\theta_L} \frac{\theta^{\theta_{obs}}e^{-\theta}}{\theta_{obs}!} d\theta= 0.9\end{equation} 
and
\begin{equation} \sum_{n=\theta_{obs}+1}^{\infty} \frac{\theta_L^n e^{-\theta_L}}{n!}=0.9\end{equation} 
give the same limit values. Frequentist and Bayesian upper limits with uniform prior are same. This is not specific to Poisson distribution and valid for all distributions where the estimated parameter is a location parameter \cite{Jaynes1976} although for Poisson distribution mean is not a location parameter. Hence this is not a trivial result for Poisson distribution.
\subsubsection{Measurement with a TS}
\label{nonideal}
The ideal Poisson counting experiment is a special case of measurement with a TS where the TS distributions of each detection count are separate. When we check the relationship between frequentist and Bayesian with uniform prior limits we see that there is not such a coincidence like the ideal counting experiment. This can be demonstrated with a simple counter example. 
Consider having uniform TS distribution for 0 detections in between [a,b] and for 1 detection in between [c,d] such that \(c<b\) and \(b-a>d-c\). TS distributions for other detection counts are separate such that there is no ambiguity there. 
For a TS measurement which may correspond to 0 or 1 count (\(c<TS_m<b\)) one needs to solve the Equation \eqref{bayeslim} in order to find the Bayesian limit for uniform prior
\begin{equation}
\begin{split}
e^{-\theta}(1+\frac{\theta(d-c)^{-1}}{(b-a)^{-1}+(d-c)^{-1}}) \\
=e^{-\theta}(1+\frac{\theta(b-a)}{b-a+d-c})=0.1
\end{split}
\label{bayeslim}
\end{equation} 
where we plugged in \(P_1(TS_m)=(d-c)^{-1}\) and \(P_0(TS_m)=(b-a)^{-1}\) to Equation \eqref{eq:bayesfinal}.

However, in the frequentist approach we solve Equation \eqref{freqlim}
\begin{equation} \label{freqlim}
e^{-\theta}(1+\frac{\theta(b-c)}{d-c})=0.1\end{equation} 
which follows from Equation \eqref{eq:freqcdf}.
In frequentist approach, the length of the intersection of \(P_0(TS)\) and \(P_1(TS)\) $(b-c)$ affects the limit whereas in the Bayesian interpretation it has no such direct role. When the intersections between the distributions vanish, we get the ideal counting example.
\section{Upper limit for fluence with a gravitational wave skymap}
\label{fluence}
Now instead of the neutrino count we want to estimate the fluence. Neutrino detectors, have a sky position dependent sensitivity due to the interaction of neutrinos and cosmic rays with the Earth and atmosphere; hence detected number of neutrinos depend on their position on the sky for a constant fluence. Again start by writing the likelihood, denote the fluence by \(\phi\),
\begin{multline} \mathcal{L}(\phi;TS_m,\mathcal{P}_{GW})=P(TS_m,\mathcal{P}_{GW}|\phi)=P(\mathcal{P}_{GW}|\phi)P(TS_m|\mathcal{P}_{GW},\phi)\\=\alpha P(TS_m|\mathcal{P}_{GW},\phi)=\alpha \sum_{n=0}^{\infty}P(TS_m|\phi,n,\mathcal{P}_{GW})P(n|\phi,\mathcal{P}_{GW})\\=\alpha \sum_{n=0}^{\infty}P(TS_m|\phi,n,\mathcal{P}_{GW})\int P(n|\phi, \Omega,\mathcal{P}_{GW})P(\Omega|\phi,\mathcal{P}_{GW})d\Omega,\ \phi\geq 0\end{multline} 
where \(n\) is the signal neutrino number from the joint source, \(\Omega\) is the sky position of source and \(\mathcal{P}_{GW}\) is the probability distribution of sky position acquired from the gravitational wave detection, namely the skymap. \(P(\Omega|\phi,\mathcal{P}_{GW})\) has no fluence dependency and is \(\mathcal{P}_{GW}(\Omega)\). \(P(\mathcal{P}_{GW}|\phi)\) doesn't have a fluence dependency and is denoted with \(\alpha\). Since it doesn't affect the maximum likelihood estimator it will be dropped for the rest of analysis. Also denote the position dependent coefficient which relates fluence to the expected neutrino number as \(c(\Omega)\) which is proportional to the effective area of the neutrino detector \cite{Aartsen_2017}.
Again assume we are sure that we only have at most 2 neutrinos, then we have

\begin{multline}\mathcal{L}(\phi;TS_m,\mathcal{P}_{GW})=P_0(TS_m)\int e^{-c(\Omega)\phi}\mathcal{P}_{GW}(\Omega)d\Omega+P_1(TS_m)\int c(\Omega)\phi e^{-c(\Omega)\phi}\mathcal{P}_{GW}(\Omega)d\Omega\\+P_2(TS_m)\int \frac{(c(\Omega)\phi)^2}{2}e^{-c(\Omega)\phi}\mathcal{P}_{GW}(\Omega)d\Omega
\label{eq:likeli}
\end{multline}

After taking the derivative with respect to \(\phi\) and equating to 0 for finding the local maximum, we have the condition for \(\phi\)

\begin{multline} \int \mathcal{P}_{GW}(\Omega)e^{-c(\Omega)\phi}c(\Omega)\\  \times[P_1(TS_m)-P_0(TS_m)+(P_2(TS_m)-P_1(TS_m))c(\Omega)\phi-P_2(TS_m)\frac{c(\Omega)^2\phi^2}{2}]d\Omega=0
\end{multline} 

Although for neutrino detectors, \(c(\Omega)\) can be well approximated; since \(\mathcal{P}_{GW}(\Omega)\) doesn't have an apriori estimated form we cannot go further in solving the equation analytically. Therefore in order to find the maximum likelihood estimator \(\hat{\phi}\) for fluence, one needs to find it numerically with known \(\mathcal{P}_{GW}(\Omega)\). 
\\
Similar to the neutrino count case, the upper limit for fluence is found by finding the \(\phi\) which satisfies Equation \eqref{phi}
\begin{equation} \label{phi}P(\hat{\phi}<\hat{\phi}'|\phi)=0.9\end{equation} 
where \(\hat{\phi}'\) is the maximum likelihood estimator for \(\phi\), and \(\hat{\phi}\) is the maximum likelihood estimator for our current measurement.
\section{Upper limit on isotropically equivalent emission energies of an ensemble of events with volume localization}
\label{sec:eiso}
In this section we consider finding an upper limit for an ensemble of similar GW events , for example same kind of events like binary black hole mergers (BBH) or binary neutron star (BNS) mergers. Due to expected different distances of these events, it can't be expected to have a similar neutrino count or fluence from each event due to the suppression with distance squared. Instead one quantity which can be similar for them is the isotropically equivalent emission energy (\(\rm{E_{iso}}\)) for neutrinos. Here by assuming all the events having the same \(\rm{E_{iso}}\), we describe the procedure of finding an upper limit on \(\rm{E_{iso}}\) with maximum likelihood estimation. Although it is clear that true \(\rm{E_{iso}}\) will be different for each event, this assumption enables us to infer more stringent information about the physics involved in same kind of processes. In order this assumption to be meaningful, the set of events should be downselected for having same kind of events. For example using BBH and BNS events together doesn't make sense as the physics involved in those are different. 

We consider having \(N\) events with volume localizations \(\mathcal{V}_{GW,i}\) and measured TS values \(TS_{m,i}\) for \(i^{th}\) event. We write the likelihood for \(\rm{E_{iso}}\) whose value is denoted as \(E_{iso}\).

\begin{equation}
\begin{split}
\mathcal{L}(&E_{iso};TS_{m,1...N},\mathcal{V}_{GW,1...N})\\=&\prod_{i=1}^{N}P(TS_{m,i},\mathcal{V}_{GW,i}|E_{iso})\\=&\prod_{i=1}^{N}P(TS_{m,i}|E_{iso},\mathcal{V}_{GW,i})P(\mathcal{V}_{GW,i}|E_{iso})=\beta\prod_{i=1}^{N}P(TS_{m,i}|E_{iso},\mathcal{V}_{GW,i})\\=&\beta\prod_{i=1}^{N}\sum_{n=0}^{\infty}P(TS_{m,i}|E_{iso},\mathcal{V}_{GW,i},n)P(n,|E_{iso},\mathcal{V}_{GW,i})\\=&\beta\prod_{i=1}^{N}\sum_{n=0}^{\infty}P(TS_{m,i}|E_{iso},\mathcal{V}_{GW,i},n)\\ &\times \int P(n|E_{iso},\mathcal{V}_{GW,i},r,\Omega)P(r,\Omega|E_{iso},\mathcal{V}_{GW,i})r^2drd\Omega, E_{iso}\geq 0
\end{split}
\end{equation}
where \(n\) is the number of signal neutrinos from the joint source, \(r\) is the distance of the source, \(\Omega\) is the sky position of the source. We assume the GW volume localizations are not affected by \(\rm{E_{iso}}\) and hence are effectively a constant for the likelihood which is denoted by \(\beta\) which will be dropped. For having at most 2 neutrinos from the joint source the likelihood becomes

\begin{multline}
\mathcal{L}(E_{iso};TS_{m,1...N},\mathcal{V}_{GW,1...N})\\=\prod_{i=1}^{N} \int (P_{0,i}(TS_{m,i})+c'(\Omega)\frac{E_{iso}}{4\pi r^2}P_{1,i}(TS_{m,i})+\frac{1}{2}(c'(\Omega)\frac{E_{iso}}{4\pi r^2})^2P_{2,i}(TS_{m,i})) \\ \times e^{-c'(\Omega) \frac{E_{iso}}{4\pi r^2}}\mathcal{V}_{GW,i}(r,\Omega)r^2drd\Omega
\end{multline}
where \(P_{0,i},\ P_{1,i},\ P_{2,i}\) are the TS distributions corresponding to 0, 1, and 2 signal neutrinos for the \(i^{th}\) event and
\begin{equation}
    c'(\Omega)=\frac{c(\Omega)}{\int E_{\nu}\frac{dN}{dE_{\nu}}dE_{\nu}}
\end{equation}
since \(\phi\) and \(\rm{E_{iso}}\) are connected as
\begin{equation}
    \int E_{\nu}\phi\frac{dN}{dE_{\nu}}dE_{\nu}=\frac{E_{iso}}{4\pi r^2}
\end{equation}
where \(E_{\nu}\) is the neutrino energy and \(\frac{dN}{dE_{\nu}}\) is the energy dependency of the differential neutrino fluence; i.e. \(=E_{\nu}^{-2}\) spectrum  over a range of energies which is expected from relativistic jet outflows \cite{1997PhRvL..78.2292W}. Similarly to the fluence case, we can't go further in analytically maximizing the likelihood. The maximum likelihood estimators should be found by numerically evaluating the likelihood  with known \(\mathcal{V}_{GW,i}(r,\Omega)\). Then the 90\% upper limit is found for \(\rm{E_{iso}}\) by finding \(E_{iso}\) which satisfies
\begin{equation}
P(\hat{E_{iso}}<\hat{E_{iso}}'|E_{iso})=0.9
\label{eiso}
\end{equation}
where \(\hat{E_{iso}}'\) is the maximum likelihood estimator for \(E_{iso}\) and \(\hat{E_{iso}}\) is the maximum likelihood estimator for the measurements of the ensemble of events.

\section{Limits for the GW events in aLIGO's first observing run}
\label{sec:res}
\begin{figure}
    \centering
    \includegraphics[width=0.5\textwidth]{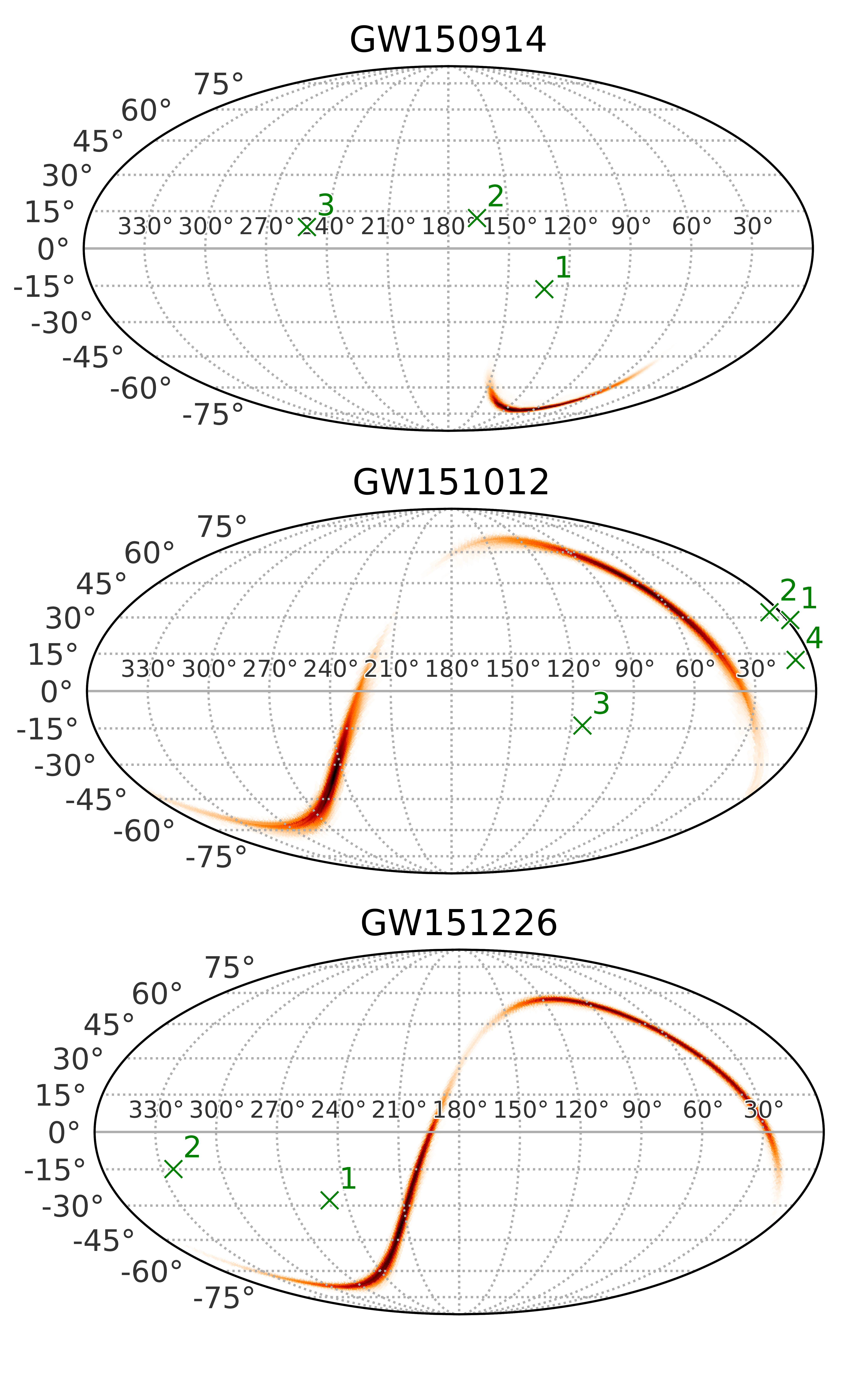}
    \caption{Sky localization \cite{parameter_estimation} of the three GW events in equatorial coordinates overlayed with neutrinos according to the labelling in Table \ref{table:neutrino}. Darker color represents higher probability density for the GW source location. Green crosses shows the location of the neutrinos.}
    \label{fig:skymaps}
\end{figure}

Now, by using publicly available data, we illustrate our method and find the neutrino emission limits on the 3 GW events from aLIGO's first observing run O1; GW150914, GW151012 and GW151226 \cite{PhysRevLett.116.241103,PhysRevX.6.041015,PhysRevLett.116.061102,Abbott_2019}. These events were analyzed before and the temporally coincident neutrinos in \(\pm500\)s window are reported \cite{PhysRevD.93.122010, PhysRevD.96.022005}. The list of neutrinos can be found in Table \ref{table:neutrino} and the GW localizations overlayed with the neurinos can be found in Fig. \ref{fig:skymaps}. Volume localization of the GW events are also available in \cite{parameter_estimation}. We use the significance calculation method of the Low-Latency Algorithm for Multi-messenger Astrophysics (LLAMA) search which has been used in the third observing run of aLIGO and aVirgo (O3) for joint GW-HEN event search \cite{PhysRevD.100.083017,countryman2019lowlatency}. 9 of the 10 reported neutrinos were detected by the IceCube Observatory and the other one was detected by ANTARES Observatory which was coincident with GW151226 \cite{PhysRevD.96.022005}. Here we only use the neutrinos detected by IceCube for simplicity as the method for calculating the significance assumes a single neutrino detector at the moment. We assume these 3 GW events are certain detections without any significance ambiguity. The significance calculation uses detector specific background distributions. In the case of certain GW events, only a background neutrino sample is needed. We use the most recent publicly available all-sky point source sample of IceCube from year 2012 in its final configuration with 86 strings\cite{public}. We assume the sensitivity of IceCube hasn't changed significantly from 2012 to 2015 as it has reached to its final configuration with 86 strings in 2011 \cite{Aartsen_2015}. 
\begin{table}
\centering
\resizebox{\textwidth}{!}{
\begin{tabular}{|l|l|l|l|l|l|l|}
\hline
Event    & \multicolumn{1}{c|}{\begin{tabular}[c]{@{}c@{}}Neutrino \\ number\end{tabular}} & Time difference $[s]$ & Right ascension $[\degree]$ & Declination $[\degree]$ & \begin{tabular}[c]{@{}l@{}}Angular \\ uncertainty {[}o{]}\end{tabular} & Energy $[TeV]$ \\ \hline
GW150914 & 1                                    & 37.2                    & 132.6                   & -16.6               & 0.35                        & 175              \\ \hline
GW150914 & 2                                    & 163.2                   & 167.0                   & 12.0                & 1.95                        & 1.22             \\ \hline
GW150914 & 3                                    & 311.4                   & -108.5                 & 8.4                 & 0.47                        & 0.33             \\ \hline
GW151012 & 1                                    & -423.3                  & 360.0                     & 28.7                & 3.5                         & 0.38             \\ \hline
GW151012 & 2                                    & -410.0                  & 7.5                     & 32.0                & 1.1                         & 0.45             \\ \hline
GW151012 & 3                                    & -89.8                   & 115.5                   & -14.0               & 0.6                         & 13.7             \\ \hline
GW151012 & 4                                    & 147.0                   & 9.0                       & 12.3                & 0.3                         & 0.35             \\ \hline
GW151226 & 1                                    & -290.9                  & 325.5                   & -15.1               & 0.1                         & 158              \\ \hline
GW151226 & 2                                    & -22.5                   & 88.5                    & 14.9                & 0.7                         & 6.3              \\ \hline
\end{tabular}}
\caption{List of neutrinos which were in the $\pm500s$ window around the three GW events. Data taken from \cite{PhysRevD.93.122010, PhysRevD.96.022005}}
\label{table:neutrino}
\end{table}
\subsection{Event generation}
The neutrino detection rate in the sample is about 3.5 in 1000s, vast majority of which we assume to be unassociated with GWs. Consequently we have the average background neutrino rate of 3.5 per GW, as we consider neutrinos in \(\pm500\)s window around the GW event. In order to construct the significance distribution for zero signal neutrinos, we drew average of 3.5 neutrinos from the public events list according to a Poisson distribution and distributed these events uniformly on a \(\pm500\)s window around the GW event. We calculated significance for 1000 such events for obtaining \(P_0\) for each event. For obtaining \(P_1\), we generated signal neutrinos by using the effective area distribution of IceCube for the same configuration which is also publicly available. First we generate the sky coordinates of these neutrinos by sampling coordinates from the sky according to the probability distribution of GW localization from the parameter estimation sample. We also keep the distance information of the chosen points from the as distance distribution is needed for \(\rm{E_{iso}}\) calculation and distance and sky position are not independent. We assume the time of the GW event is not a random variable and therefore all the events with the same GW data create the same localization. The effective area depends on the neutrino energy and the declination of neutrinos\cite{Aartsen_2017}. By assuming an \(E_{\nu}^{-2}\) spectrum \cite{2013JCAP...06..008A,2013RvMP...85.1401A,2014PhRvD..90j2002A,PhysRevD.93.122010,Albert_2017,PhysRevD.96.022005,Albert_2019} and using the effective area, we assign these neutrinos energies. However the data sample has the reconstructed energies which can be different than the true energy \cite{public}. By using the distribution in Fig. 20 of \cite{Aartsen_2014} we convert the true energies to reconstructed energies. This energy variation is not taken in account in the significance calculation method, therefore it could lead unphysical variations in the calculated limits if the significance TS was used directly. By generating TS distributions for various scenarios (i.e $P_0$, $P_1$) the suggested method in this study minimizes such effects when these effects were not considered in the previous stages of the analysis. One remaining property of the generated neutrinos is their angular uncertainty. In order not to lose the dependency of angular uncertainty to energy and declination, for each generated neutrino we narrowed the list of real events via their energies and declinations. We required declination difference with the generated neutrino to be less than \(10\degree\) and differences of their base 10 logarithm of energies to be less than 1.5. From the remaining neutrinos we picked one of their angular uncertainty uniformly and assigned it to the generated neutrino. We set the energy and declination difference constraints by considering the number of neutrinos remained after cuts. Having more stringent cuts causes some of the generated neutrinos to have no similar real neutrino for taking its angular uncertainty. With a larger sample, more stringent cuts could be imposed. Due to the axial symmetry of IceCube we did on put a constraint on the right ascension of neutrinos. Next, we shift the position of each generated neutrino, by randomly choosing a offset distance according a normal distribution with zero mean and with variance angular error squared. Then a uniformly random angle is chosen in \(0-2\pi\) and neutrinos' positions on the sky is shifted along that direction by the chosen distances. After the neutrinos are generated we sampled the time of each signal neutrino from a symmetric triangular distribution whose mode is the GW event time and extend is \(\pm500\)s. This distribution is obtained if one assumes the GW and neutrino emission to be uniformly in \(\pm250\)s window around the same astrophysical event \cite{BARET20111}. Convolution of two uniform distributions give a triangular distribution which implies that temporally closer neutrinos to the GW are more likely to be associated than background neutrinos which are distributed uniformly around the GW event. Finally we choose background neutrinos to accompany each signal neutrino. These background neutrinos are chosen identically as the ones chosen for obtaining \(P_0\). Similarly we generate 1000 of such events. We do not consider any of these neutrinos to be coming from the same source as for each combination of two, the positional difference is larger than the sum of angular errors. Hence we take \(P_2\) to be zero for all of the events.
\subsection{Calculation and results}
\begin{figure}
    \centering
    \includegraphics[width=0.5\textwidth]{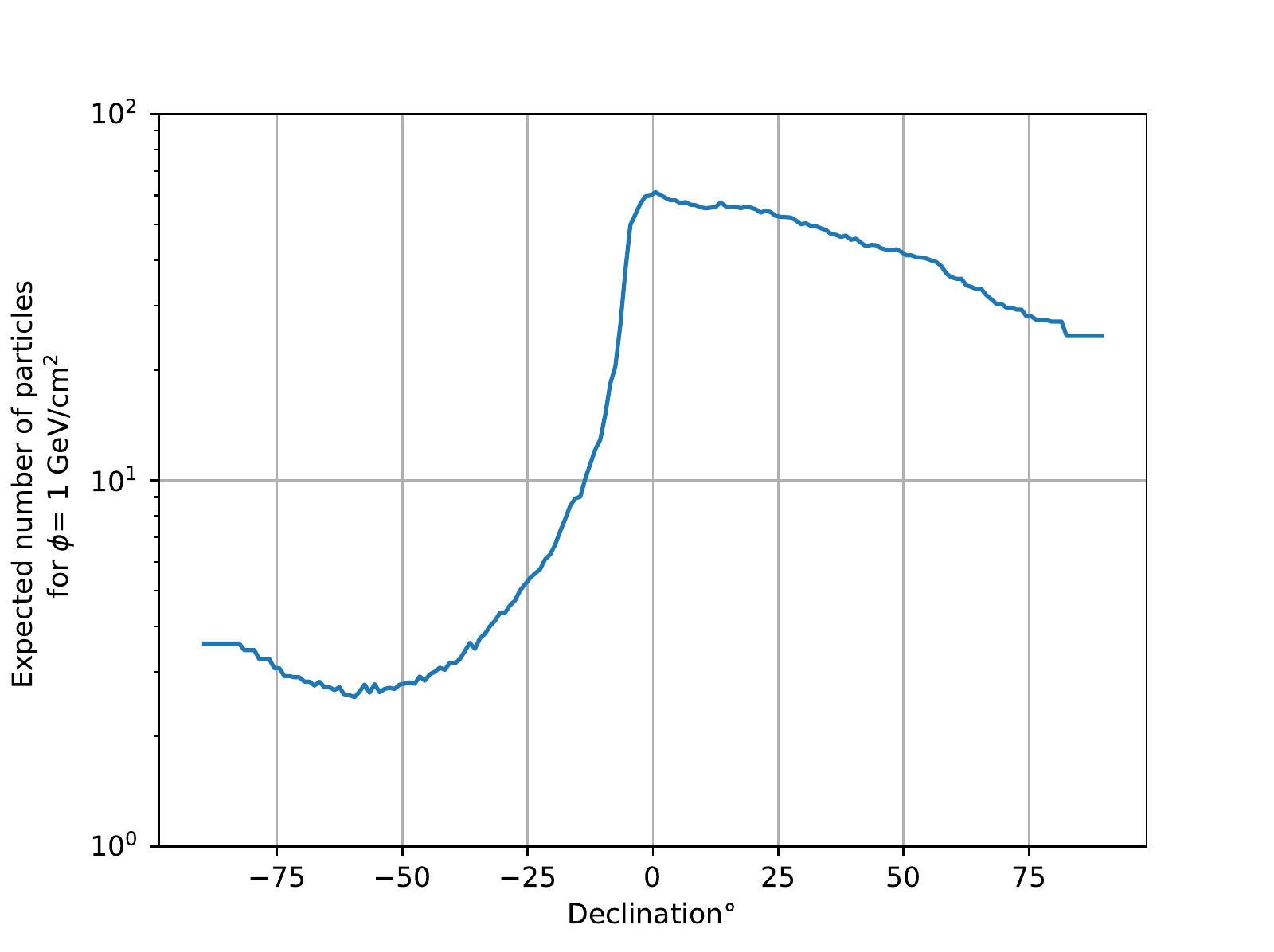}
    \caption{Expected number of particles for the differential fluence of \(1\  GeV/cm^2 E_{\nu}^{-2}\) vs. declination, obtained from the effective area distribution}
    \label{fig:c}
\end{figure}

We find the maximum likelihood estimators for the events with actual coincident neutrinos, as well as the generated background and signal neutrinos. Then we find the value of the true quantity which satisfies Eqs. \eqref{theta}, \eqref{phi} or \eqref{eiso}. In order to do it, and also for finding the estimators for fluence and \(\rm{E_{iso}}\), we need to find the expected number of neutrinos for unit fluence at a given declination which has been denoted as \(c(\Omega)\) in this paper. The requirement when calculating the value of the true quantity arises due to the fact that we directly sampled the signal neutrino positions from the GW skymap; but didn't account for the declination dependency of the IceCube's sensitivity or the effective area. Up to this point we only used the effective area for the energies of the neutrinos after we had chosen the positions. By using \(c(\Omega)\) we get the expected signal neutrino count distribution for the unit fluence for every point in the sky, which assures the use of full effective area dependency in the calculation. With that distribution we weight the signal neutrino events with the Poisson probabilities whose mean is determined by the true value of the estimated quantity, declination and the distance of the simulated emission. For events with 2 or more signal neutrinos we take the corresponding estimator to be higher than the estimator of the actual event. \(c(\Omega)\) is obtained by integrating the effective area of each declination in energy after scaling with \(E_{\nu}^{-2}\). For \(E_{\nu}^{-2}\) spectrum the obtained \(c(\Omega)\) is shown in  Fig \ref{fig:c}.

The obtained 90\% frequentist upper limits with maximum likelihood estimators are shown in Table \ref{table}. The Bayesian upper limits for the neutrino count with uniform prior are found to be 2.3 for all events.

\begin{table}
\centering
\begin{tabular}{|l|l|l|l|}
\hline
             Event& \multicolumn{1}{c|}{\begin{tabular}[c]{@{}c@{}} Signal \\ neutrino count\end{tabular}} & \begin{tabular}[c]{@{}l@{}}Fluence \\ $[GeV/cm^2]$\end{tabular} & $\rm{E_{iso}}$ $[ergs]$ \\ \hline
GW150914     &             2.3                                                                            &       0.80                                                           &   $5.1\times10^{53}$   \\ \hline
GW151012     &      2.2                                                                                   &     0.40                                                             &  $1.1\times10^{54}$    \\ \hline
GW151226     &    2.3                                                                                     &    0.45                                                              &   $2.6\times10^{53}$   \\ \hline
Combined &       -                                                                                  &        -                                                          &  $6.1\times10^{52}$    \\ \hline
\end{tabular}
\caption{90\% upper limits for the signal neutrino count, fluence and $\rm{E_{iso}}$ for three events and $\rm{E_{iso}}$ for the population of three events. The combined limit is found by following the description in Section \ref{sec:eiso} and by assuming the same emission energy for all events.}
\label{table}
\end{table}

\subsection{Discussion}
The 90\% frequentist upper limit for the mean count of a background free ideal Poisson process is 2.3 and as shown in Sec. \ref{sec:bayescomp} it also corresponds to a 90\% Bayesian upper limit with uniform prior for the count. When we look at the neutrino count limits we have we see that except the frequentist limit for the event GW151012 we have the upper limit as 2.3 neutrinos. It shows that for GW150914 and GW151226 the background and signal distributions are separated sufficiently for behaving as ideal counting processes. The fact that for GW151012, the frequentist limit is less than 2.3 points a substantial intersection of background (\(P_0\)) and signal (\(P_1\)) distributions in a region of higher significance than the event's significance. The fact that its Bayesian limit is still 2.3 implies that for the values around event's significance the intersection of background and signal distributions is negligible. When we look at the fluence upper limits, we see that the upper limit for GW150914 is about twice of other events' limits. This can be explained by the localization of the GWs and IceCube's declination dependent sensitivity, i.e. Fig. \ref{fig:c}. Required fluence for the same number of neutrinos is about an order larger in the south hemisphere compared to the northern hemisphere, therefore fluence requirement from southern regions dominates the northern regions for comparable probabilities. As GW150914 is completely localized in the southern hemisphere and the other two events are more or less equally localized in both hemispheres, it is expected to have twice the limits of GW151012 and GW151226 for GW150914. When we look at the upper limits for \(\rm{E_{iso}}\), we see that the limit of GW150914 is twice of GW151226. Both events have similar median distances, 440 Mpc and 450 Mpc for GW150914 and GW151226 respectively \cite{Abbott_2019}. Therefore we expect their \(\rm{E_{iso}}\) upper limit ratio to be similar to their fluence upper limits. The expected distance of GW151012 is 1080 Mpc \cite{Abbott_2019}, about twice of GW150914 and GW151226. Therefore we expect a factor of 4 difference between the fluence upper limits and  \(\rm{E_{iso}}\) upper limits, which is present. In \cite{PhysRevD.93.122010, PhysRevD.96.022005} \(\rm{E_{iso}}\) upper limits were found for every point in the sky. Our method allows one to have a single upper limit value for the whole event. Our upper limits fall in the previously reported upper limit range in the whole sky. Finally we comment on the \(\rm{E_{iso}}\) upper limits for combination of the three events. We see that when three GW events which don't have sufficient significance for having a counterpart in neutrino emission are combined for an upper limit, we get an order of magnitude more stringent upper limit. This illustrates the importance of having collection of events when constraining astrophysical parameters.

\section{Conclusion}
We described the methods of finding frequentist upper limits with maximum likelihood estimators for expected neutrino count, neutrino emission fluence and isotropically equivalent neutrino emission energy for an ensemble of events which are specifically aimed for joint GW and HEN events; but could be used for any similar search which uses TS for counting discrete events. Then we applied this method on the GW events in aLIGO's first observing run (O1) and found upper limits for them.
Through the paper we considered 90\% upper limits although different confidence intervals at arbitrary confidence levels can also be found with this method instead of 90\% upper limits; by requiring different relationships between the estimators of the true quantity and the measurement instead of Equations \eqref{theta}, \eqref{phi} or \eqref{eiso}.
\section{Acknowledgments}
The authors are grateful for the useful discussion with Klas Hulqvist and Hans Niederhausen, and comments from Austin Schneider. The authors thank Columbia University in the City of New York and University of Florida for their generous support. The Columbia Experimental Gravity group is grateful for the generous support of the National Science Foundation under grant PHY-1708028. DV is grateful to the Ph.D. grant of the Fulbright foreign student program.

\bibliographystyle{JHEP}
\bibliography{upbib}

\end{document}